# Fast-ion conduction and flexibility of glassy networks

**Deassy I. Novita, P. Boolchand**,
Department of Electrical and Computer Engineering, University of Cincinnati,
Cincinnati, OH 45221-0030, USA

**M.Malki,**
Centre de Recherche sur les Materiaux a Haute Temperature, CNRS 4212, 1D Av. de la Recherche Scientifique, 45071 Orleans, and Polytech'Orléans – Université d'Orléans, 8, rue Léonard de Vinci, 45072 Orléans, France

**M. Micoulaut**
Laboratoire de Physique Theorique de la Matiere Condensée, University Pierre et Marie Curie, Boite 121, 4, Place Jussieu, 75252 Paris, Cedex05, France.

We observe two thresholds in the variations of electrical conductivity of *dry* solid electrolyte $(AgI)_x(AgPO_3)_{1-x}$ glasses, when the AgI additive concentration x increases to 9.5% and to 37.8%. Raman scattering complemented by calorimetric measurements confirm that these thresholds are signatures of the rigidity phase transitions; at x = 9.5% from a stressed rigid to an isostatically (stress free) rigid phase, and at x = 37.8% from isostatically rigid to a flexible phase. In the flexible phase, the electrical conductivity seems to increase as a power of x, this is in good agreement with the theoretical prediction based on 3d percolation.

The solid electrolytes, AgI, $Ag_2S$, $Ag_2Se$, exist in a non-crystalline or glassy phase, usually not as stoichiometric solids but as additives in base network glasses [1]. These additives can either segregate [1, 2] as separate phases, or uniformly mix [1] with the base glass to form homogeneous solid electrolyte glasses. Gaining a more complete understanding of ion-transport in these systems is a basic scientific challenge, with important technological consequences. These materials find use in batteries, sensors, non-volatile memories for portable devices [3] and electro-chromic displays [4]. The solid electrolyte $(AgI)_x(AgPO_3)_{1-x}$ appears to form homogeneous glasses, and their physical behavior [5-7] including compositional variations in electrical conductivity [8, 9] has been examined rather extensively although a consensus on the data has been elusive. Here we show that the variability of the data is likely due to water contamination. Our work is on specifically prepared *dry* samples, and reveals physical properties display *two* thresholds, one near x = 9.5% and a second near x = 37.8%. We show that these thresholds separate different elastic regimes in network structure.

Phase diagrams of disordered solids based on connectedness of their backbones have their origin in the simple and elegant ideas of mechanical constraints. Notion of constraints in mechanics was introduced by J.L.Lagrange [10], and applied to understand mechanical stability of macroscopic structures by J.C. Maxwell [11], and to model elastic behaviour of covalent glassy networks by J.C. Phillips [12] and M.F. Thorpe [13]. That there are actually three (Flexible, Intermediate and Stressed-Rigid) and not two (Flexible and rigid) elastic phases of disordered solids is a more recent development [14-17] in the field that has opened new avenues to understand, the unfolding process of proteins [13], design of thin-film gate dielectrics for transistors [15], Satisfiability problems in Computational science [16], and the near absence of aging of covalent glassy networks [17] in Intermediate Phases. Our observations here bring electrolyte glasses under the same generic classification [18, 19], and highlight network *flexibility* to be the functionality that promotes fast-ion-conduction.

AgPO$_3$ possesses a glass a transition temperature [20], (T$_g$) of 254°C, when synthesized by handling usual precursors in a *dry* (R.H. < 0.2%) ambient atmosphere. But T$_g$s of the glass decrease to the 160-190°C range when precursors are handled in laboratory ambient environment (RH ~ 50%) [8, 9, 20]. The role of bonded water in lowering T$_g$ of a AgPO$_3$ glass was recognized earlier [21], although the highest T$_g$ realized in the earlier work was only 189°C. Present solid electrolyte glass samples were synthesized by weighing and intimately mixing 99.999% Ag$_3$PO$_4$, P$_2$O$_5$, and AgI as fine powders in a dry N$_2$ gas purged glove box [20] (R.H.< 0.2%), and reacting them at 900°C. Melts were equilibrated at 600°C, and quenched on steel plates, and glass samples cycled through T$_g$ to relieve frozen stress [20]. A model 2920 modulated-DSC from TA Instruments operated at 3°C/min scan rate and 1°C/100s modulation rate was used to examine glass transitions [17]. A Solartron SI 1260 Impedance spectrometer was used to study AC electrical conductivity [18] as a function of temperature in the 200K < T < T$_g$ range, and frequency in the $1 < f < 10^6$ Hz range. Here we report on room temperature f→0 (dc) conductivity results. Raman scattering was excited using 514 nm radiation, and the scattered radiation analysed using a model T64000 triple monochromater system [17].

The observed variations in T$_g$(x) and the non-reversing enthalpy, ΔH$_{nr}$(x), of present *dry* (AgI)$_x$(AgPO$_3$)$_{1-x}$ glasses are summarised in Fig.1(a) and Fig.1(b). Here x represents the mole fraction of AgI. We find T$_g$(x) to monotonically decrease as AgI content increases, but the ΔH$_{nr}$(x) term to vary non-monotonically displaying a rather striking global minimum (reversibility window [17]) in the 9.5% < x < 37.8% range. At higher x (> 45%), the ΔH$_{nr}$(x) term decreases again as glasses depolymerise. Variations in room temperature electrical conductivity, σ(x), appear in Fig.1(c), and show increases in steps, one near 9.5% and another near 37.8%. For comparison, we have shown in Fig.1(a) and 1(c), variations in $T_g(x)$ and in $\sigma(x)$ reported by earlier groups [9, 22, 23]. The present findings on dry samples differ significantly from previous ones in the field.

The thermal and electrical results above lead to the obvious question- are there vibrational anomalies associated with structure of these electrolyte glasses as were observed earlier in covalent systems [17, 24]? This, indeed, is the case as revealed by our Raman scattering results (Fig.2). The base glass is widely believed [20, 25, 26] to consist of chains of quasi-tetrahedral $PO_4$ units with each P atoms having two bridging ($O_b$) and two terminal ($O_t$) oxygen near-neighbors. In the base glass (x = 0), modes near 1140 cm$^{-1}$ and 684 cm$^{-1}$ are identified [20, 25, 26] with symmetric vibrations of P-$O_t$ and P-$O_b$ of these tetrahedra in polymeric chains. The asymmetric counterparts of these modes are weakly excited in Raman but strongly in IR [20, 25, 26]. And with increasing x, these modes steadily red-shift and decrease in scattering strength as two new pairs of modes, one near 1000 cm$^{-1}$ and 750 cm$^{-1}$, and a second one near 960 cm$^{-1}$ and 720 cm$^{-1}$ appear, and steadily grow in scattering strength. The first and second pair of modes are identified [26] respectively with $PO_4$ tetrahedra present in small-rings and large-rings (Fig.2). These results lead to a picture of these glasses as being chain-like at low-x (< 30%) but becoming ring-like at high-x (> 50%), features that are in harmony with decreasing molar volumes [6, 27] and a loss of network connectivity independently supported by reduction of $T_g$s within the agglomeration theory [28]. Raman and IR [20] vibrational density of states change with glass composition, and these results are in contrast to earlier reports [23, 27] that reveal little or no change.

The observed Raman lineshapes when analyzed in terms of a superposition of Gaussians provide variations in frequency of the P-$O_t$ symmetric mode near 1140 cm$^{-1}$ (Fig. 2 and 3a). The mode is found to steadily red-shift displaying two vibrational thresholds, one near x = 9.5%, and the other near x = 37.8% which correlate rather well with the walls of the reversibility window (fig.1b) and the steps in electrical conductivity. Red-shift of the mode in question occurs as the inter-chain spacings increase due to insertion of AgI lowering the global connectivity of the backbone, a feature that has parallels in covalent glasses [17, 24]. The underlying optical elasticity varies with glass composition as a power-law, which is deduced by plotting the square of Raman mode frequency ($v^2 - v_c(1)^2$) against glass composition ($x - x_c$), and in the $0 < x < 9.5\%$ range,

$$v^2 - v_c(1)^2 = A\,(x - x_c(1))^{p_1} \qquad (1)$$

and yields (Fig.3b) the power-law $p_1 = 1.25(2)$. Here $v_c(1)$ represents the value of $v$ at $x = x_c(1) = 9.5(3)\%$. The power-law was deduced by plotting the log ($v^2 - v_c(1)^2$) against the log ($x - x_c(1)$), with the slope of the line yielding $p_1$. In the $9.5\% < x < 37.8\%$ range, the corresponding power-law is found (Fig.3c) to be $p_2 = 0.98(3)$ with $x_c(2) = 37.8(5)\%$. As noted earlier, room temperature conductivities increase with x (Fig.1c) to display a change in regime near $x_c(1) = 9.5\%$, and near $x_c(2) = 37.8\%$. If we fit the increase of conductivity at $x > x_c(2)$ to a power-law,

$$\sigma(x) = B\,(x - x_c(2))^{\mu} \qquad (2)$$

we obtain a value of $\mu = 1.78(10)$ and of $x_c(2) = 37.8(5)\%$ as shown in Fig. 1c and Fig. 3d.

The thermal, optical and electrical results presented above lend themselves to the following interpretation. The reversibility window, $9.5\% < x < 37.8\%$, in analogy to the case of covalent glasses [17, 24], we identify with the *Intermediate phase* of the present solid electrolyte glasses. The base AgPO$_3$ glass is weakly *stressed-rigid*, and AgI alloying steadily lowers the connectivity of the chain network as reflected in the reduction of T$_g$(x) and emergence of the first sharp diffraction peak [5] near 0.7 A$^{-1}$ in neutron scattering experiments. The Raman optical elastic power-law of p$_1$ = 1.25 (2), for glasses in the 0 < x < 9.5% range, is in reasonable accord with numerical predictions [29] for the power-law in stressed-rigid networks ( p$^{theo}$ = 1.4), and the observed values of the elasticity power-law in the stressed-rigid covalent glasses [17, 24]. Taken together, the results show that glasses in the 0 < x < 9.5% range possess backbones that are *stressed-rigid*, and that the threshold composition, x$_c$(1) = 9.5%, represents the stress-transition [14]. Currently, there are no theoretical estimates for the elastic power-law in Intermediate Phases but we note that the present value of p$_2$ = 0.98(3) is in excellent agreement with the value observed in covalent glasses [17, 24]. Thus, the reversibility window, Raman optical elastic thresholds, and elastic power-laws, show that glasses in the 0 < x < 9.5% range possess backbones that are *rigid but mildly stressed*, in the 9.5% < x < 37.8% range these are *rigid but stress-free,* and in the 37.8% < x < 55% range these are elastically *flexible*. At higher x (> 55%), glasses segregate into AgI-rich regions, as observed T$_g$s acquire values characteristic of AgI glass [1].

Addition of the electrolyte salt AgI to the insulating base AgPO$_3$ glass serves to provide Ag$^+$ carriers, and to also elastically soften the base glass. At low x (< 9.5%), Ag$^+$ ions undergo *localized* displacements in backbones as suggested by Reverse Monte Carlo simulations [5], a view that is independently corroborated by intrinsically *stressed-rigid* character of these glasses in the present work. With increasing AgI, and particularly in the Intermediate Phase, backbones become stress-free [24] and Ag$^+$ displacements increase as do conductivities (Fig.1c). At higher x (> 37.8%), backbones become elastically flexible and electrical conductivities increase precipitously as carriers freely diffuse [5] along percolative pathways. Thus, although carrier concentrations increase monotonically with x, the observed thresholds in σ(x) suggest that it is network rigidity (flexibility) that controls fast-ion conduction by suppressing (promoting) $Ag^+$ ion migration.

The increase of ionic conductivity upon an elastic softening of a glass network represents an example of a complex system in which one functionality (elasticity) affects another (conductivity). The conductivity power-law in the present glasses (μ = 1.78(10)), and in $(K_2O)_x(SiO_2)_{1-x}$ glasses (μ = 1.77) [18] when they become flexible, is in reasonable agreement with an electronic conductivity power-law (μ= 2.0) predicted in 3D bond depleted resistor networks at the percolation (connectivity) threshold [30, 31]. In summary, properties of flexibility and rigidity of glassy networks commonplace in covalent systems [14, 17, 24], apparently extend to solid electrolyte glasses as well, and fast ion-conduction is promoted when networks become flexible. We thank Professor B. Goodman and Professor D. McDaniel for continued discussions on glasses. This work is supported in part by the NSF grant DMR 04-56472.

.

**Figure 1.** Variations in (a) $T_g(x)$, (b) non-reversing enthalpy $\Delta H_{nr}(x)$ (c) room temperature conductivities, $\sigma(x)$ in dry $(AgI)_x(AgPO_3)_{1-x}$ glasses synthesized in present work (▲) and those reported by Mangion-Johari (●) ref.[22] and Sidebottom (●) ref. [9] and Bhattacharya et al (●) ref.[23]. At x > 55%, $T_g$s decrease to near 65° C, a value characteristic of AgI glass (see ref. [1].The reversibility window in $\Delta H_{nr}(x)$ fixes the Intermediate Phase, as in covalent glasses (see ref. [17, 24]).

**Figure 2.** Raman lineshapes in *dry* $(AgPO_3)_{1-X}(AgI)_X$ glasses show vibrational modes as follows: $a_1$ (684.7 cm$^{-1}$) and $a_4$(917.4 cm$^{-1}$ ) represent symmetric and asymmetric modes of P-O$_{br}$ in long-chains, $a_2$(722.8 cm$^{-1}$) & $a_5$ (967 cm$^{-1}$) represent symmetric and asymmetric modes of P-O$_{br}$ in large-rings, $a_3$ (761.9 cm$^{-1}$) & $a_6$(1004.6 cm$^{-1}$) represent symmetric and asymmetric modes of (P-O$_{br}$) in small-rings, $a_7$ represent an asymmetric mode of PO$_3^{2-}$ species (Q$_1$ species) at 1095.1 cm$^{-1}$ , and $a_8$ (1140 cm$^{-1}$) & $a_9$(1245.4 cm$^{-1}$) represent symmetric and asymmetric stretch modes of P-O$_t$ in long-chains. Glasses transform from a chain-like to ring-like as the AgI content increases. The red-shift of mode $a_8$ with x is analysed in Fig.3.

**Figure 3.** (a) Raman $a_8$ mode frequency red-shifts with increasing x to display two thresholds, one near $x_c(1)$ = 9.5% and a second near $x_c(2)$ = 37.8%. (b) shows a plot of log ( $v^2 - v_c(1)^2$) against log ($x_c(1) - x$) in the 0 < x < 9.5% range, and gives a slope $p_1$ = 1.25(2). (c) shows a plot of log ($v^2 - v_c(2)^2$) against log ($x_c(2) - x$) in the 9.5% < x < 37.8% range and gives a slope $p_2$ = 0.98(3).(d) shows a plot of log $\sigma$ against log (x – $x_c(2)$) and yields a conductivity power-law, $\mu$ = 1.78(10) in the flexible phase with $x_c(2)$ = 37.8%.

10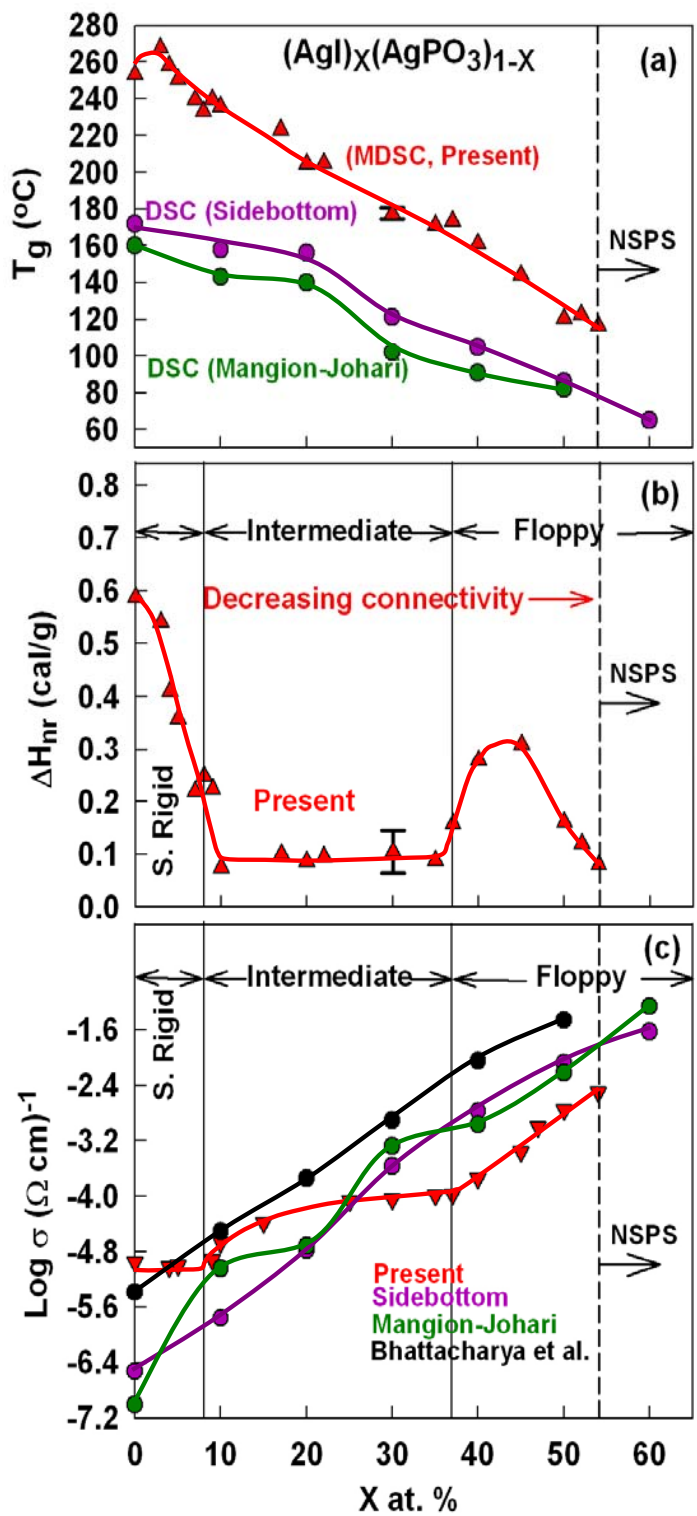

Figure 1

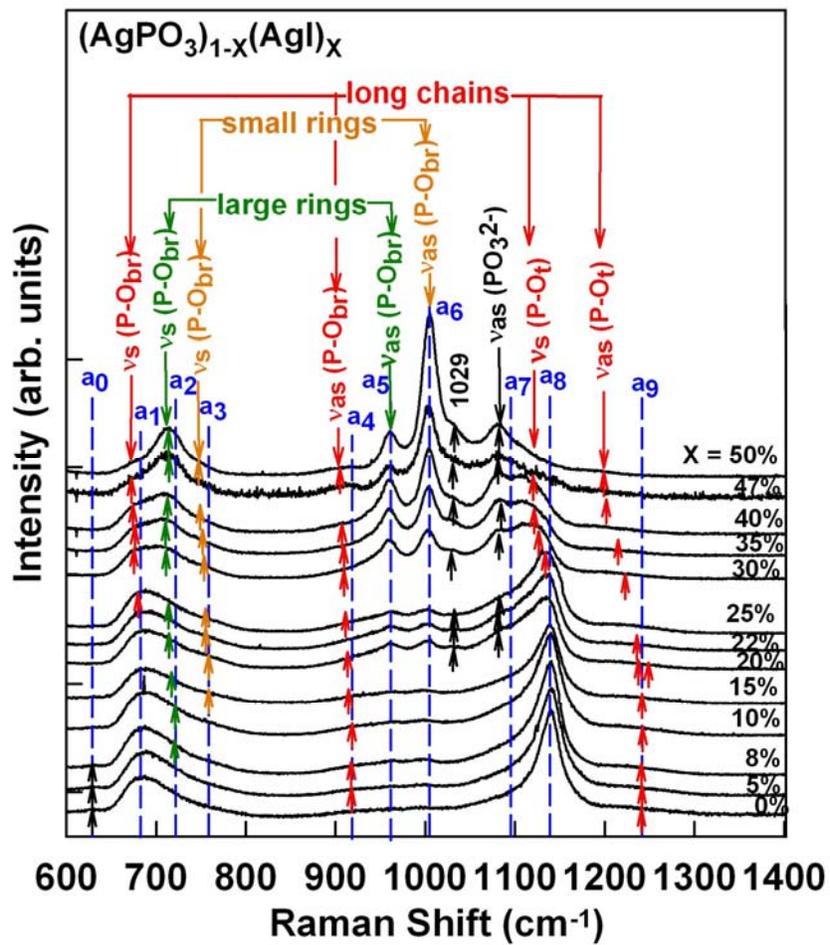

Figure 2

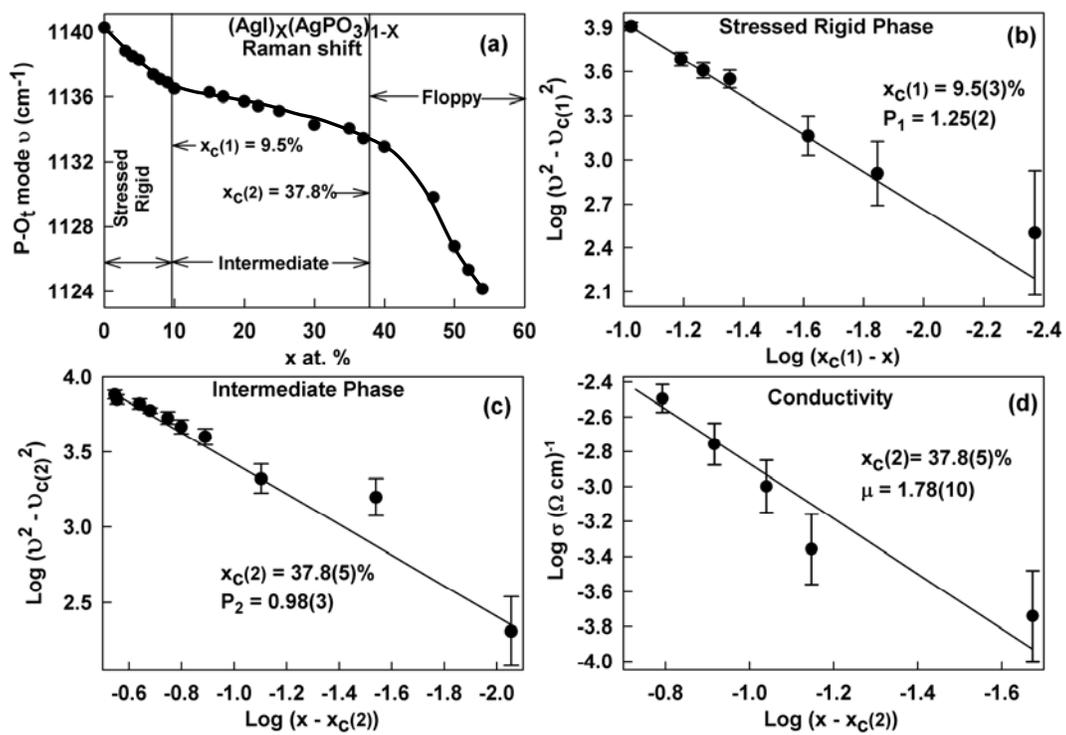

Figure 3